\documentclass[preprint,showpacs,superscriptaddress,prd]{revtex4}
\usepackage{graphicx,amsmath,dcolumn,bm}
\begin{document}
\title{\Large{\bf{Nuclear geometry effect and transport coefficient in  semi-inclusive lepton-production of hadrons
 off nuclei }}}
\author{Na  Liu }
\email[E-mail: ]{Liuna@sjzue.edu.cn}

\affiliation{Department of Physics, Hebei Normal
             University,
             Shijiazhuang 050024, P.R.China}
\affiliation{College of Mathematics and Physics, Shijiazhuang
University of Economics, Shijiazhuang 050031, P.R.China}

\author{Wen-Dan  Miao }
\email[E-mail: ]{miaowd@mail.hebtu.edu.cn}

\affiliation{Department of Physics, Hebei Normal
             University,
             Shijiazhuang 050024, P.R.China}

\author{Li-Hua Song }
\email[E-mail: ]{songlh@mail.heuu.edu.cn}

\affiliation{Department of Physics, Hebei Normal
             University,
             Shijiazhuang 050024, P.R.China}
\affiliation{ College of Science,  Hebei United University, Tangshan
063009, P.R.China}

\author{Chun-Gui  Duan }
\email[E-mail: ]{duancg@mail.hebtu.edu.cn}

\affiliation{Department of Physics, Hebei Normal
             University,
             Shijiazhuang 050024, P.R.China}




\begin{abstract}

Hadron production in semi-inclusive deep-inelastic scattering of
leptons from nuclei is an ideal tool to determine and constrain the
transport coefficient  in cold nuclear matter. The leading-order
computations for hadron multiplicity ratios are performed by means
of the SW  quenching weights and the analytic parameterizations of
quenching weights based on BDMPS formalism. The theoretical results
are compared to the HERMES positively  charged pions production data
with the quarks hadronization occurring outside the nucleus. With
considering the nuclear geometry effect on hadron production, our
predictions are in good agreement with the experimental
measurements.   The extracted  transport parameter from the global
fit is shown to be $\hat{q} = 0.74\pm0.03$ GeV$^2$/fm for the SW
quenching weight without the finite energy corrections. As for the
analytic parameterization of BDMPS quenching weight without the quark
energy $E$ dependence, the computed transport coefficient is
$\hat{q} = 0.20\pm0.02$ GeV$^2$/fm.  It is found that the nuclear
geometry effect has a significant impact on the transport
coefficient in cold nuclear matter. It is necessary to consider the
detailed nuclear geometry in studying the semi-inclusive hadron
production in deep inelastic scattering on nuclear targets.

\vskip 0.1cm

\noindent{\bf Keywords:}  deep inelastic scattering, quark energy
loss, hadron production.

\end{abstract}

\pacs{ 25.30.-c;  
       24.85.+p ; 
       13.60.Le ; 
       12.38.-t; 
       }

\maketitle
\newpage
\vskip 0.5cm

\section{Introduction}
Hadron production in semi-inclusive deep inelastic scattering of
leptons from nuclei is a good tool to understand the parton
propagation and hadronization processes in cold nuclear matter. In
the semi-inclusive deep inelastic scattering on nuclei, a virtual
photon from the incident lepton is absorbed by a quark within a
nucleus, the highly virtual colored quark propagates over some
distance through the cold nuclear medium,  evolves subsequently into
an observed hadron. It is expected that the hadron multiplicities
observed in the scattering is sensitive to whether the hadronization
occurs within or outside the nucleus in a semi-classical picture.
Therefore, the detailed understanding of the parton propagation and
hadronization processes in cold nuclear matter would greatly benefit
the study of the jet-quenching and parton energy loss phenomena
observed in ultra-relativistic heavy-ion collisions [1].

Hadron production in deep-inelastic scattering was first performed
at SLAC [2] followed by EMC [3], E665 [4]. The more precise
experimental measurements were reported by the HERMES [5-8] and CLAS
[9,26]. The experimental data are usually presented in terms of the
multiplicity ratio $R^h_M$, which is defined as the ratio of the
number of hadrons $h$ produced per deep-inelastic scattering  event
on a nuclear target with mass number $A$ to that for a deuterium
target. However, the experimental results on $R^h_M$ were only a
function of one kinematic variable  (so-called one-dimensional
dependences) except that a two-dimensional dependence was reported
for a combined sample of charged pions [8].  Recently, the hadron
multiplicities in semi-inclusive deep-inelastic scattering were
measured on neon, krypton, and xenon targets relative to deuterium
at an electron(positron)-beam energy of 27.6 GeV at HERMES [10]. The
multiplicity ratios were measured in a two-dimensional
representation for positively and negatively charged pions and
kaons, as well as protons and antiprotons. The two-dimensional
representation consists in a fine binning in one variable and a
coarser binning in another variable. The published two-dimensional
data will further help us to study the space (time) development of
hadronization in more detail.

In our preceding articles, we have calculated the nuclear
modifications of hadron production in semi-inclusive deep inelastic
scattering  in a parton energy loss model. By means of the short
hadron formation time, the relevant data with quark hadronization
occurring outside the nucleus are picked out from HERMES
experimental results [8] on the one-dimensional dependence of the
multiplicity ratio $R^h_M$ as a function of the energy fraction $z$
of the virtual photon carried away by the hadron. Our theoretical
results show that the nuclear effects on parton distribution
functions can be neglected. It is found that the theoretical results
considering the nuclear modification of fragmentation functions due
to quark energy loss are in good agreement with the selected
experimental data.  The obtained energy loss per unit length is
$0.38 \pm 0.03$ GeV/fm for an outgoing quark by the global fit [11].
Furthermore, the leading-order computations on hadron multiplicity
ratios for the production of positively and negatively charged pions
and kaons are compared with the selected two-dimensional data from
the HERMES experiment [10] by means of the hadron formation time[8,
27], which is estimated from the Lund model. For the case where the
quark hadronization occurs inside the nucleus [12], it is shown that
with increase of the energy fraction carried by the hadron, the
nuclear suppression on hadron multiplicity ratio from nuclear
absorption gets bigger. The nuclear absorption is the dominant
mechanism causing a reduction of the hadron yield. The atomic mass
dependence of hadron attenuation  is confirmed theoretically and
experimentally to be proportional to  $A^{1/3}$. For the case where
the quark hadronization occurs outside the nucleus [13], the atomic
mass number dependence is  theoretically and experimentally in good
agreement with the $A^{2/3}$ power law. It should be noted that in
these studies, the outgoing quark energy loss is employed in the
form of the mean energy loss. The length traveled by the quark in
the nuclear matter is taken as the average path length  $L=(3/4)R_A$
where $R_A$ is the nuclear radius.

In this paper, we employ  the two-dimensional data from HERMES on
the multiplicity ratios for positively  charged pions production on
neon nucleus  with respect to deuterium target [10], seeing that the
CLAS data[9,26] are suitable for studying the hadron production
inside the nuclei. The experimental data with quark hadronization
occurring outside the nucleus are selected by means of the  hadron
formation time in order to investigate the quark energy loss in cold
nuclear matter. Because the mean energy loss can not reflect the
dominant parton energy loss mechanism, the present study will use
the Salgado-Wiedemann(SW) quenching weights [14], and the analytic
parameterizations of quenching weights[28] based on Baier,
Dokshitzer, Mueller, Peign$\acute{e}$ and Schiff (BDMPS)
formalism[22]. In addition, the nuclear geometry effect on the
extracted transport coefficient is investigated. It is hoped to gain
new knowledge about quark energy loss in a cold nuclear medium.

The remainder of the paper is organized as follows. In Section II,
the brief formalism for the hadron  multiplicity in semi-inclusive
deep inelastic scattering on the nucleus and nuclear modification of
the fragmentation functions owing to quark energy loss are
described. In Section III, The results and discussion obtained are
presented. Finally, a summary is presented.

\section{The hadron  multiplicity in semi-inclusive deep inelastic
scattering on nuclei }

Hadron production in semi-inclusive deep inelastic scattering can be
computed  within the QCD improved parton model. At leading order in
perturbative QCD, the hadron multiplicity is given by
\begin{equation}
\frac{1}{N^{DIS}_{A}}\frac{dN^{h}_{A}}{dzd
\nu}=\frac{1}{\sigma^{lA}}\int
dx\sum_{f}e^{2}_{f}q^{A}_{f}(x,Q^{2})\frac{d\sigma^{lq}}{dxd\nu}D^{A}_{f|
h}(z,Q^{2}),
\end{equation}
\begin{equation}
\sigma^{lA}=\int
dx\sum_{f}e^{2}_{f}q^{A}_{f}(x,Q^{2})\frac{d\sigma^{lq}}{dxd\nu},
\end{equation}
where $N_{A}^{h}$ ($N_{A}^{DIS}$) is the yield of semi-inclusive
(inclusive) deep-inelastic scattering leptons on nuclei $A$, $\nu$
is the virtual photon energy, $e_f$ is the charge of the quark with
flavor $f$, $q^{A}_{f}(x,Q^{2})$ is the nuclear quark distribution
function with Bjorken variable $x$ and photon virtuality $Q^{2}$,
$D^{A}_{f| h}(z,Q^{2})$ is the nuclear modified fragmentation
function of a quark of flavour $f$ into a hadron $h$. The
differential cross section for lepton-quark scattering at leading
order, ${d\sigma^{lq}}/{dxd\nu}$, is given by
\begin{equation}
\frac{d\sigma^{lq}}{dxd\nu}=Mx\frac{4\pi\alpha_{em}^{2}}{Q^{4}}[1+(1-y)^{2}],
\end{equation}
where $M$ and  $ \alpha _{em}$ are respectively the nucleon mass and
the fine structure constant, $y$ is the fraction of the incident
lepton energy transferred to the target.

In semi-inclusive deep inelastic scattering, the initially produced
energetic quark will have to go through multiple scattering and
induced gluon bremsstrahlung as it propagates through the medium.
The induced gluon bremsstrahlung effectively reduces the leading
parton energy.  The quark energy fragmenting into a hadron shifts
from $E=\nu$ to $E'=\nu-\Delta E$, which results in a rescaling of
the energy fraction of the produced hadron:
\begin{equation}
z=\frac{E_{h}}{\nu}  \longrightarrow  z'=\frac{E_{h}}{\nu-\Delta E},
\end{equation}
where $E_{h}$ and $\Delta E$ are respectively the measured hadron
energy and the quark energy loss in the nuclear medium. The
fragmentation function in the nuclear medium[15],
\begin{equation}
D^{A}_{f|h}(z,Q^{2})= \int_{0}^{(1-z)\nu}d(\Delta E) P(\Delta
E,\omega_{c},L)\frac{1}{1-\Delta E/\nu}D_{f|h}(z',Q^{2}),
\end{equation}
where $D_{f|h}$ is the standard (vacuum) fragmentation function of a
quark of flavour $f$ into a hadron $h$, the characteristic gluon
frequency $\omega_{c}=(1/2)\hat{q}L^{2}$,  $P(\Delta
E,\omega_{c},L)$ denotes the probability for a quark with energy $E$
to lose an energy $\Delta E$ which originates from the quark
radiating gluons. This probability distribution, the so-called
quenching weight,  depends on the path length $L$ covered by the
hard quark in the nuclear medium and the transport coefficient
$\hat{q}$.

Based on the above formalism, the path length $L$ is fixed, only the
transport coefficient $\hat{q}$ remains to be determined.

\section{Results and discussion}

In order to explore the outgoing quark energy loss, we will pick out
the experimental data with quark hadronization outside the nucleus
from the two-dimensional data on the multiplicity ratio $R^h_M$ for
positively  charged pions production on neon nucleus in three $z$
slices as a function of $\nu$, and in five $\nu$ slices as a
function of $z$ [10].  Unlike our previous researches with using the
mean path length  $L=(3/4)R_A$, we set $t>2R_A$ with the hadron
formation time $t=z^{0.35}(1-z)\nu/\kappa$ ($\kappa = 1GeV/fm$)
[8,11,27], which is defined as the time between the moment that the
quark is struck by the virtual photon and the moment that the
prehadron is formed. This assumption can make sure that the hadron
is produced outside target nucleus rather than inside the nucleus.
Using the criterion, the number of points in selected experimental
data is 4 for $R^{{\pi}^+}_{M}(\nu)$ for  positively charged  pions
produced on neon target in the $z$ region of $0.2< z < 0.4$. For the
multiplicity ratio $R^{{\pi}^+}_{M}$ as a function of $z$, there are
15 data points in three $\nu$ regions of 14 $< \nu <$ 17 GeV , 17 $<
\nu <$ 20 GeV and 20 $< \nu <$ 23.5 GeV. In total, our analysis has
19 data points.

To comparing to the selected experimental data,  we compute at
leading order  the hadron multiplicity ratios $R^{{\pi}^+}_{M}$,
\begin{equation}
R^{{\pi}^+}_{M}[\nu(z)]=\int
\frac{1}{N^{DIS}_{A}}\frac{dN^{h}_{A}(\nu,z)}{dzd\nu}dz(\nu)
\Bigg/\int
\frac{1}{N^{DIS}_{D}}\frac{dN^{h}_{D}(\nu,z)}{dzd\nu}dz(\nu).
\end{equation}
In our calculation,  the vacuum fragmentation functions [17] is
employed together with the CTEQ6L parton density in the proton [16].
By means of the CERN subroutine MINUIT [18], the transport
coefficient $\hat{q}$ is obtained by minimizing $\chi^2$. One
standard deviation of the optimum parameter  corresponds to an
increase of $\chi^{2}$ by 1 unit from its minimum $\chi^{2}_{min}$.


\begin{table}[t,m,b]
\caption{The values of  $\hat{q}$ and  $\chi^{2}/ndf$ extracted from
the selected data on  the hadron multiplicity ratios for positively
charged pions produced on neon target by means of the SW quenching
weights [14]. }
\begin{ruledtabular}
\begin{tabular*}{\hsize}
{c@{\extracolsep{0ptplus1fil}} c@{\extracolsep{0ptplus1fil}}
c@{\extracolsep{0ptplus1fil}} c@{\extracolsep{0ptplus1fil}}
c@{\extracolsep{0ptplus1fil}}}
          & \multicolumn{2}{c}{non-reweighted}& \multicolumn{2}{c}{reweighted}
          \\ \cline{2-3} \cline{4-5}
           & $\hat{q}(GeV^{2}/fm)$ &$\chi^{2}/ndf$& $\hat{q}(GeV^{2}/fm)$ &$\chi^{2}/ndf$\\
         \hline
         & \multicolumn{4}{c}{$L=(3/4)R_{A}$} \\
               \cline{2-5}
        $0.2<z<0.4$ & $1.30\pm0.06 $ & $0.02 $& $1.44\pm0.09$ & $0.009$\\
       $14<\nu<17$ &$1.29\pm0.05$ &$0.13$& $1.44\pm0.02$& $0.19$\\
       $17<\nu<20$  & $1.10\pm0.09 $& $0.53 $& $1.14\pm0.06$& $0.58$\\
       $20<\nu<23.5$ &$0.94\pm0.06$&$0.62$& $0.94\pm0.01$ & $0.61$\\
       Global fit &$1.15\pm0.03$&$0.47$& $1.20\pm0.02$ & $0.54$\\
       \hline
          & \multicolumn{4}{c}{$L=\sqrt{R_{A}^{2}-b^{2}}-y$} \\ \cline{2-5}
        $0.2<z<0.4$ & $0.84\pm0.02$ & $0.013 $& $1.02\pm0.06$ & $0.006$\\
       $14<\nu<17$ &$0.83\pm0.04$ &$0.05$& $1.07\pm0.02$& $0.13$\\
       $17<\nu<20$  & $0.71\pm0.05 $& $0.37 $& $0.79\pm0.05$& $0.47$\\
       $20<\nu<23.5$ &$0.60\pm0.05$&$0.53$& $0.60\pm0.02$ & $0.55$\\
       Global fit &$0.74\pm0.03$&$0.36$& $0.83\pm0.02$ & $0.46$\\
\end{tabular*}
\end{ruledtabular}
\end{table}

We use the SW quenching weight[14]in the soft multiple scattering
approximation with the mean path length of the outgoing quark in the
nuclear medium $L=(3/4)R_A$. In this case, the determined transport
coefficient $\hat{q}$ and $\chi^2$ per number of degrees of freedom
($\chi^2/ndf$) are summarized in Table I for positively charged
pions production on neon nuclei  in a $z$ slice  and in three $\nu$
slices  from the HERMES experiments[10]. The solid curves in Fig. 1
are our numeric results which are compared with the selected
experimental data. It is shown that the computed hadron multiplicity
ratio increases with $\nu$, and decreases with $z$.  It is found
that the theoretical results are in good agreement with the
experimental data.  The global fit of all data makes $\hat{q}
=1.15\pm0.03 $ GeV$^2$/fm with the relative uncertainty $\delta
\hat{q} / \hat{q} \simeq 2.6\%$ and $\chi^2/ndf = 0.47$.

\begin{figure}[t,m,b]
\centering
\includegraphics*[width=15cm,height=12cm]{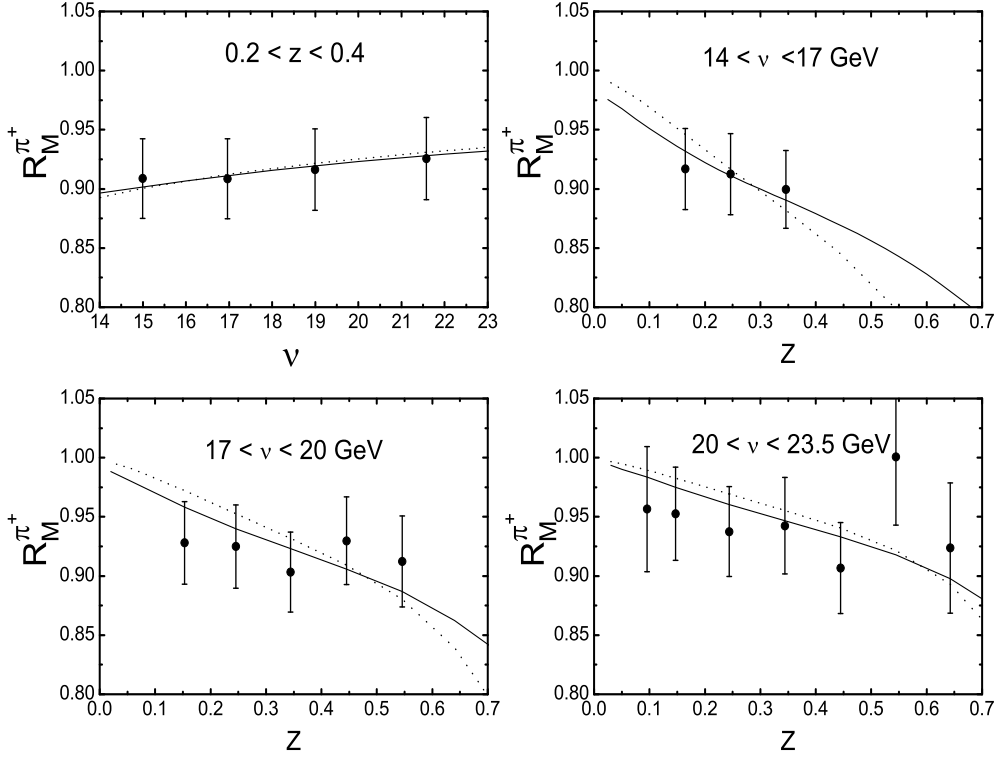}
\vspace{-1.1cm} \caption{The calculated multiplicity ratios
$R^{{\pi}^+}_{M}$ in one $\nu$ slice and three $z$ slices for
positively charged pions production on neon target from the
non-reweighted SW quenching weight(solid curves) and the
parameterization of BDMPS quenching weight for the quark energy E
independence (dotted curves). The HERMES data [10]  are shown with
the total uncertainty (statistical plus systematic, added
quadratically).}
\end{figure}

In above calculation, the employed quenching weight is computed in
the eikonal limit of very large quark energy [14].  For the case of
realistic initial quark energy,  the finite energy correction need
to be taken into account. To illustrate the theoretical uncertainty
associated with finite energy corrections, we shall compare results
without such corrections to those  from the reweighted  probability
[19]
\begin{equation}
P_{rw}(\Delta E/\omega_{c},L)=\frac{P(\Delta
E/\omega_{c},L)}{\int_{0}^{\nu/\omega_{c}}d(\Delta
E/\omega_{c})P(\Delta E/\omega_{c},L)}\Theta (1-\Delta E/E).
\end{equation}
Regarding the  reweighted quenching weight, the global fit of all
data gives $\hat{q} =1.20\pm0.02$ GeV$^2$/fm with the relative
uncertainty $\delta \hat{q} / \hat{q} \simeq 1.6\%$ and $\chi^2/ndf
= 0.54$(see Table I).  It is clear that the determined transport
coefficient $\hat{q}$ is about  $4\%$ greater than that from
non-reweighted quenching weight.

\begin{figure}[t,m,b]
\centering
\includegraphics*[width=10cm,height=6.0cm]{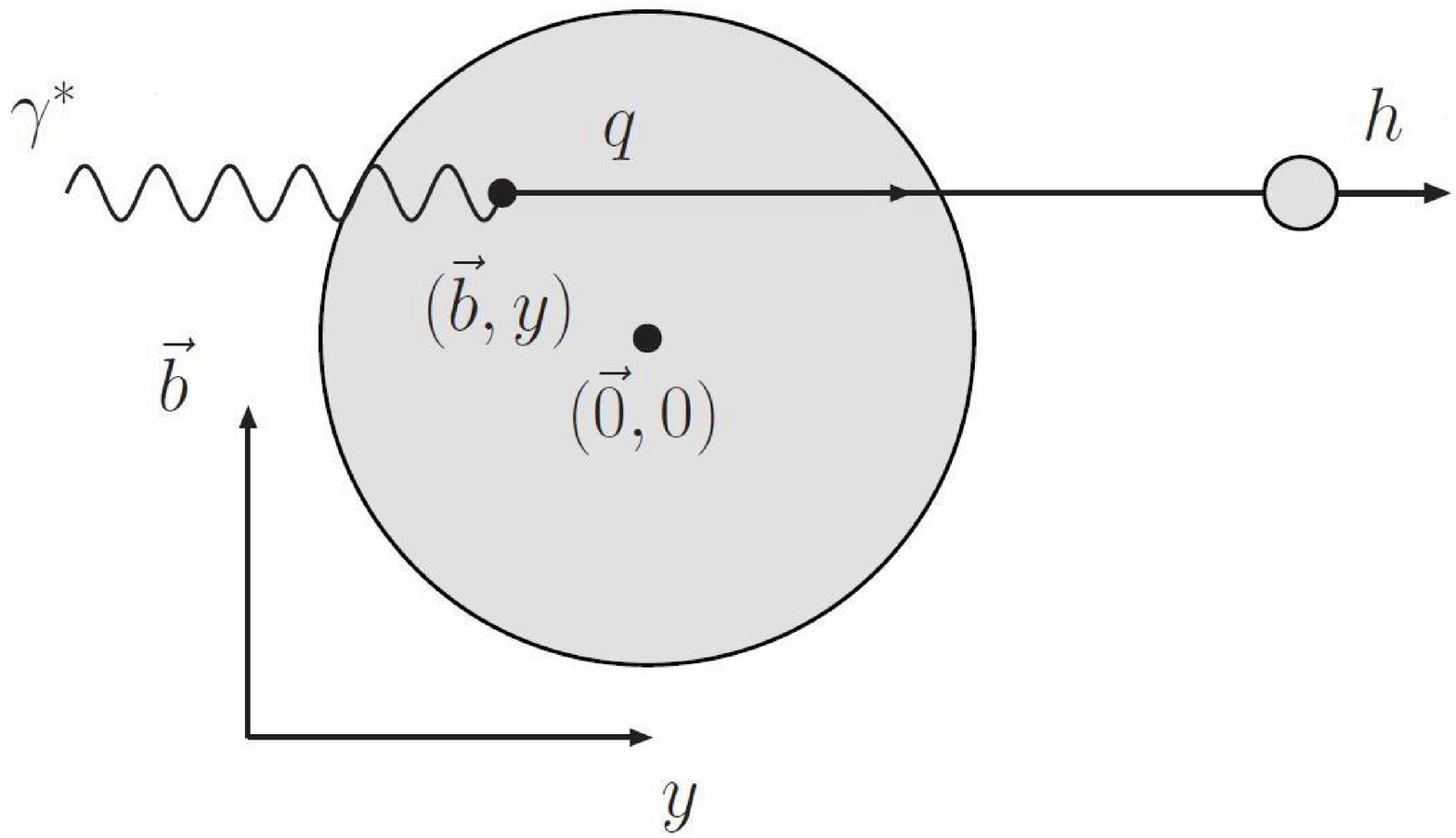}
\vspace{-1.0cm} \caption{Illustration of the geometry of hadron
production in deeply inelastic scattering in the nucleus target rest
frame. See details in text.}
\end{figure}

Now let us to study the nuclear geometry effect in the
semi-inclusive lepto-production of hadrons off nuclei. The nuclear
geometry is described by the nuclear density distribution
$\rho_A(\vec{b}, y)$, where $y$ is the coordinate along the
direction of the outgoing quark and $\vec{b}$ its impact parameter.
The center of the target nucleus lies at $(\vec{0}, 0)$ (see Fig.
2). If we assume that the interaction of the virtual photon  with
the quark is located  at $(\vec{b}, y)$,  the colored quark produced
at $y$ will travel the path length $L=\sqrt{R_{A}^{2}-b^{2}}-y$,
along a direction with impact parameter $\vec{b}$. Then the averaged
modified fragmentation function [20,27,29] should be
\begin{equation}
D^{A}_{f|h}(z,Q^{2})= \int
d^{2}bdy\rho_{A}(\vec{b},y)\int_{0}^{(1-z)\nu}d(\Delta E) P(\Delta
E,\omega_{c},L)\frac{1}{1-\Delta E/\nu}D_{f|h}(z',Q^{2}).
\end{equation}

For simplicity, we use the uniform hard-sphere nuclear density
normalized to unity,
$\rho_A(\sqrt{b^2+y^2})=(\rho_0/A)\Theta(R_A-\sqrt{b^2+y^2})$ with
$\rho_0$ is the nuclear density, $R_A=r_0{A^{1/3}}$, and $r_0 =
1.12$ fm. By combining Eq.8 and Eq.1, the calculated corresponding
transport coefficient $\hat{q}$ and $\chi^2$ per number of degrees
of freedom ($\chi^2/ndf$) are summarized in Table I with the
reweighted  and non-reweighted  quenching weights. It is shown that
the global fit of all data gives separately $\hat{q} = 0.74\pm0.03$
GeV$^2$/fm with the relative uncertainty $\delta \hat{q} / \hat{q}
\simeq 4.0\%$ and $\chi^2/ndf = 0.36$  for the non-reweighted
quenching weight, and $\hat{q} = 0.83\pm0.02$ GeV$^2$/fm with the
relative uncertainty $\delta \hat{q} / \hat{q} \simeq 2.4\%$ and
$\chi^2/ndf = 0.46$  for the reweighted quenching weight. It is
obvious that the nuclear geometry effect could reduce the transport
coefficient by as much as $36\%$ for the non-reweighted quenching
weight and $31\%$ for the reweighted quenching weight, respectively.

As can be seen from the Table I, the values of $\hat{q}$ are too
high to be understandable in perturbation theory. The reason is that
the coefficient of the discrete part $p_0\delta(\Delta E)$ of the
quenching weight,  $p_0$, is very large when using the finite-length
medium in the SW quenching weights. The calculated value of $p_0$
reveals that $p_0$($\hat{q}=0.8$ GeV$^{2}$/fm) reduces nonlinearly
from 1.07 to 0.2 in the range $1.0 \leq L \leq 6.0$
fm($2R_{Ne}\simeq 6.0$ fm). As for the same $L$, $p_0$ decreases as
the increase of $\hat{q}$. Therefore,  the small-$L$ SW quenching
weights result in the large values for $\hat{q}$.

The SW quenching weight evolves from the BDMPS calculation to
include finite-length effects. The original BDMPS calculation takes
a limit in which the nuclear medium length goes to infinity[22]. The
 analytic parameterization was given for BDMPS quenching weight[28].
The quenching weight $P(\Delta E)$ for the quark energy $E$
independence  follows with a great accuracy a log-normal
distribution,
\begin{equation}
\bar{P}(\bar{\Delta E}=\Delta E/\omega_c)=\omega_c P(\Delta E) =
\frac{1}{\sqrt{2\,\pi}\, \sigma\,\bar{\Delta
E}}\,\exp\left[-\frac{\left(\log{\bar{\Delta
E}}-\mu\right)^2}{2\,\sigma^2}\right],
\end{equation}
with $\mu=-1.5$ and $\sigma=0.73$. The quenching weight
$P(\bar{\Delta E},\bar{E})$ for the quark energy dependence follows
\begin{equation}
\bar{P}(\bar{\Delta E}, \bar{E}) = \frac{1}{\sqrt{2\,\pi}\,
\sigma(\bar{E})\,\bar{\Delta
E}}\,\exp\left[-\frac{\left(\log{\bar{\Delta
E}}-\mu(\bar{E})\right)^2}{2\,\sigma(\bar{E})^2}\right]
\end{equation}
where the $\mu(\bar{E})$ and $\sigma(\bar{E})$ are given by the
empirical laws
\begin{equation}
\begin{split}
\mu(\bar{E}) & = -1.5 + 0.81 \times \left( \exp \left( -0.2/ \bar{E} \right) -1 \right),\\
\sigma(\bar{E}) & = \,\,0.72 + 0.33 \times \left(\exp\left(-0.2/
\bar{E} \right) -1 \right).
\end{split}
\end{equation}
The analytic parameterization $P(\bar{\Delta E},\bar{E})$ reproduces
qualitatively the characteristics of the $E$ dependence of the mean
energy loss $<\Delta E>$ from BDMPS formalism[30], e.g.,  the mean
energy loss $<\Delta E>\propto \sqrt{\hat{q}E}L$  at small quark
energy($E<\omega_c$).


\begin{table}[t,m,b]
\caption{The values of  $\hat{q}$ and  $\chi^{2}/ndf$ extracted from
the selected data on  the hadron multiplicity ratios for positively
charged pions produced on neon target by means of the analytic
parameterizations of quenching weights[28]. }
\begin{ruledtabular}
\begin{tabular*}{\hsize}
{c@{\extracolsep{0ptplus1fil}} c@{\extracolsep{0ptplus1fil}}
c@{\extracolsep{0ptplus1fil}} c@{\extracolsep{0ptplus1fil}}
c@{\extracolsep{0ptplus1fil}}}
          & \multicolumn{2}{c}{$E$ independence}& \multicolumn{2}{c}{$E$ dependence}
          \\ \cline{2-3} \cline{4-5}
           & $\hat{q}(GeV^{2}/fm)$ &$\chi^{2}/ndf$& $\hat{q}(GeV^{2}/fm)$ &$\chi^{2}/ndf$\\
         \hline
         & \multicolumn{4}{c}{$L=(3/4)R_{A}$} \\
               \cline{2-5}
        $0.2<z<0.4$ & $0.38\pm0.07 $ & $0.04 $& $0.41\pm0.09$ & $0.03$\\
       $14<\nu<17$ &$0.37\pm0.07$ &$0.51$& $0.40\pm0.09$& $0.52$\\
       $17<\nu<20$  & $0.26\pm0.05 $& $0.98 $& $0.27\pm0.06$& $1.00$\\
       $20<\nu<23.5$ &$0.20\pm0.06$&$0.79$& $0.21\pm0.07$ & $0.79$\\
       Global fit &$0.29\pm0.03$&$0.83$& $0.30\pm0.04$ & $0.84$\\
       \hline
          & \multicolumn{4}{c}{$L=\sqrt{R_{A}^{2}-b^{2}}-y$} \\ \cline{2-5}
        $0.2<z<0.4$ & $0.26\pm0.05$ & $0.04 $& $0.30\pm0.06$ & $0.02$\\
       $14<\nu<17$ &$0.25\pm0.05$ &$0.33$& $0.29\pm0.07$& $0.38$\\
       $17<\nu<20$  & $0.19\pm0.04 $& $0.84 $& $0.20\pm0.05$& $0.88$\\
       $20<\nu<23.5$ &$0.15\pm0.04$&$0.74$& $0.15\pm0.04$ & $0.76$\\
       Global fit &$0.20\pm0.02$&$0.70$& $0.22\pm0.02$ & $0.74$\\
\end{tabular*}
\end{ruledtabular}
\end{table}

Taking advantage of the analytic parameterizations of quenching
weights without and with the quark energy $E$ dependence, the
computed  transport coefficient $\hat{q}$ and $\chi^2$ per number of
degrees of freedom ($\chi^2/ndf$) are listed in Table II. The
expected multiplicity ratios $R^{{\pi}^+}_{M}$ (dotted  curves) with
$L=(3/4)R_{A}$ for the quenching weight of quark energy $E$
independence, are in good agreement with the selected experimental
data in Fig.1. It is shown in Table II that the smaller values of
$\hat{q}$ are given with comparing the SW quenching weights.  The
obtained value of $\hat{q}$ should be much more realistic from the
BDMPS quenching weights.

As can be shown from the Table II, the global fit of all data by the
 quark energy $E$ independent quenching
weight    gives separately $\hat{q} = 0.29\pm0.03$ GeV$^2$/fm with
$\chi^2/ndf = 0.83$  from the mean length, and $\hat{q} =
0.20\pm0.02$ GeV$^2$/fm with  $\chi^2/ndf = 0.70$  from the nuclear
geometry effect. For the quark energy $E$ independent and dependent
quenching weight, the values of $\hat{q}$ from the global fit are
consistent in $1\sigma$ range with(without)  the nuclear geometry
effect. Moreover, the nuclear geometry effect could reduce the
transport coefficient by as much as $31\%$ and $27\%$  respectively.
Therefore, we should consider the nuclear geometry effect over the
course of studying the semi-inclusive hadron production in deep
inelastic scattering on nuclear targets.

What we notice is in the Table II that the extracted values of
$\hat{q}$ are consistent within $1\sigma$ range by fitting the
selected experimental data in the $z$ region of $0.2< z < 0.4$, and
in two $\nu$ regions of 14 $< \nu <$ 17 GeV and 17 $< \nu <$ 20 GeV.
However, in  20 $< \nu <$ 23.5 GeV region, the smaller value of
$\hat{q}$ are given  in comparison with those in other three
kinematic ranges. The reason is that in  20 $< \nu <$ 23.5 GeV
region, the value of $R^{{\pi}^+}_{M}(z=0.54)$ deviates normal trend
obviously(see Fig.1). If we remove this  experimental data point,
the calculated values of $\hat{q}$ are respectively $0.18\pm0.05$
GeV$^2$/fm ($\chi^2/ndf=0.42$, $E$ independent quenching weight) and
$0.19\pm0.06$ GeV$^2$/fm ($\chi^2/ndf=0.43$, $E$ dependent quenching
weight) with considering the nuclear geometry effect. Then the
obtained values of $\hat{q}$ can be consistent within $1\sigma$
range for the collected experimental data in four kinematic regions.
Hence, the precise experimental data can refine our theoretical
model.


\begin{table}[t,m,b]
\caption{The values of  $\hat{q}$ and  $\chi^{2}/ndf$ extracted from
the selected data($2t>2R_A$) and the full data on  the hadron
multiplicity ratios for positively charged pions produced on neon
target by considering the nuclear geometry
effect($L=\sqrt{R_{A}^{2}-b^{2}}-y$) and using the analytic
parameterization of BDMPS quenching weight for the quark energy $E$
independence[28]. }
\begin{ruledtabular}
\begin{tabular*}{\hsize}
{c@{\extracolsep{0ptplus1fil}} c@{\extracolsep{0ptplus1fil}}
c@{\extracolsep{0ptplus1fil}} c@{\extracolsep{0ptplus1fil}}
c@{\extracolsep{0ptplus1fil}}}
          & \multicolumn{2}{c}{$2t>2R_A$}& \multicolumn{2}{c}{full data}
          \\ \cline{2-3} \cline{4-5}
           & $\hat{q}(GeV^{2}/fm)$ &$\chi^{2}/ndf$& $\hat{q}(GeV^{2}/fm)$ &$\chi^{2}/ndf$\\
        $0.2<z<0.4$ & $0.22\pm0.02$ & $0.24 $& $0.22\pm0.02$ & $0.24$\\
       $14<\nu<17$ &$0.12\pm0.02$ &$2.70$& $0.06\pm0.01$& $3.98$\\
       $17<\nu<20$ & $0.10\pm0.03 $& $1.77 $& $0.10\pm0.02$& $1.59$\\
       $20<\nu<23.5$ &$0.13\pm0.04$&$0.76$& $0.13\pm0.04$ & $0.76$\\
       Global fit &$0.146\pm0.01$&$1.40$& $0.109\pm0.01$ & $2.59$\\
\end{tabular*}
\end{ruledtabular}
\end{table}

In this work, the hadron formation time, $t=z^{0.35}(1-z)\nu/\kappa$
, is quoted from HERMES Collaboration[8]. This idea was pursued in
Ref. [27]. The convenient parametrization can give its values as a
function of $z$ closely resembling the ones obtained in the Lund
model[31]. However, the kinematic dependence for the formation time
as well as its magnitude is not clearly known. To investigate the
sensitivity of $\hat{q}$ values on the  hadron formation time, our
analysis is repeated  by multiplying the formation time by 2 and 1/2
in order to select more or less data. It is found that there is no
experimental data on  $R^{{\pi}^+}_{M}$ that meet the criteria of
$(1/2)t>2R_A$ for positively charged  pions produced on neon target
in the $z$ region of $0.2< z < 0.4$, and in three $\nu$ regions of
14 $< \nu <$ 17 GeV , 17 $< \nu <$ 20 GeV and 20 $< \nu <$ 23.5 GeV.
By using $2t>2R_A$, 30 data points are selected in total for the
corresponding four kinematic regions(see Fig.3). By considering the
nuclear geometry effect and using the analytic parameterization of
BDMPS quenching weight without the quark energy $E$ dependence,  the
computed transport coefficient $\hat{q}$ and $\chi^2/ndf$ are
summarized in Table III. For completeness, meanwhile, the values of
$\hat{q}$ and $\chi^2/ndf$ are given in Table III by fitting the
full experimental data (34 data points, see Fig.3) in four kinematic
regions. The calculated results are compared with the  experimental
data on multiplicity ratios in Fig.3. The dotted and dashed lines
are the results on multiplicity ratios by fitting the selected
data($2t>2R_A$) and the full data, respectively. For convenient for
comparing, the calculated multiplicity ratios are presented in
Fig.3(solid lines) by using the selected data($t>2R_A$).

\begin{figure}[t,m,b]
\centering
\includegraphics*[width=15cm,height=12.0cm]{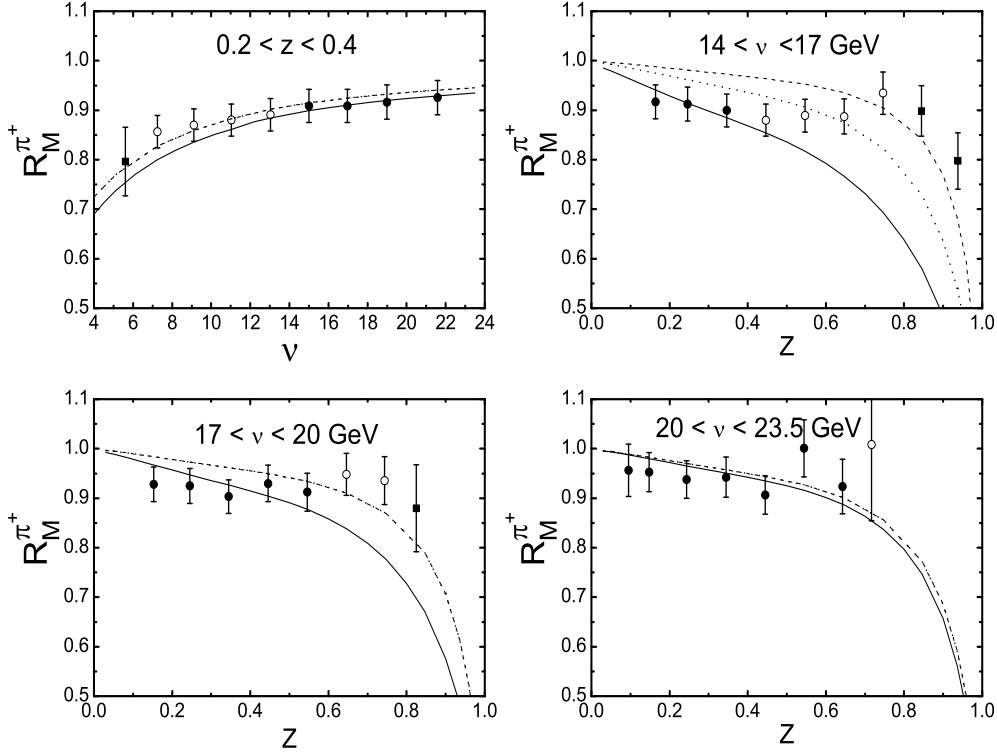}
\vspace{-1.0cm} \caption{The multiplicity ratios $R^{{\pi}^+}_{M}$.
the solid, dotted and dashed lines correspond to the computed
results by fitting  the selected data($t>2R_A$, filled circles), the
collected data($2t>2R_A$, filled  and empty circles) and the full
data, respectively.}
\end{figure}

As can be seen from the Table III and Fig.3,  the extracted values
of $\hat{q}$ are identical except the region  14 $< \nu <$ 17 GeV
for the selected data($2t>2R_A$) and the full data. In the range 14
$< \nu <$ 17 GeV,  the theoretical values of $\chi^2/ndf$ are
2.70($2t>2R_A$) and 3.98(full data) separately. Moreover, the
extracted value of $\hat{q}$ from the selected data($2t>2R_A$) is
bigger than that from the full data. For the selected
data($t>2R_A$), the collected data($2t>2R_A$) and the full data, as
a whole, the values of $\hat{q}$ by the global fit are $0.20\pm0.02
$ GeV$^2$/fm($\chi^2/ndf=0.70$), $0.146\pm0.01 $
GeV$^2$/fm($\chi^2/ndf=1.40$) and $0.109\pm0.01 $
GeV$^2$/fm($\chi^2/ndf=2.59$), respectively. As the included
experimental data point increases, the extent of hadron absorption
contamination become bigger. Additionally,  the degree of the
accordance with experimental data become worse. Although this
assumption $t>2R_A$ results in a too drastic data selection, the
selection criteria can make sure that the hadron is produced outside
target nucleus. Up to now, the formation time is yet an open
question. In order to learn about the quark energy loss, on the one
hand the in-depth study of the formation time is necessary  in
theory, but on the other the much higher energy experiments are
expected to isolate the parton energy loss  from any possible
pre-hadron absorption effects.

It is worthy of note that our obtained transport coefficients  are
larger than Accardi's fit of $\hat{q} = 0.5$ GeV$^2$/fm[20, 29]  and
Dupr$\acute{e}$'s fit of $\hat{q} = 0.4$ GeV$^2$/fm[26]   at the
center of a target nucleus  by using SW quenching weight. As for the
quenching weights based on BDMPS formalism, our results are bigger
than Arleo's value of $\hat{q} = 0.14$ GeV$^2$/fm [21]. In addition,
Deng and Wang applied the modified DGLAP evolution to quark
propagation in the deep inelastic scattering of a large nucleus
within the framework of generalized factorization for higher-twist
contributions to multiple parton scattering. The extracted jet
transport parameter at the center of a large nucleus is found to be
$\hat{q} = 0.024\pm0.008$ GeV$^2$/fm [23]. However, these research
did not include $\chi^2$ analysis, and did not distinguish whether
the hadronization occurs within or outside the nucleus for the
experimental data.

It is worth stressing that BDMPS evaluated the parton energy loss in
 the multiple-soft scattering approximation, and took a limit in
which the medium length goes to infinity. Salgado and Wiedemann
improved the BDMPS calculation to include finite-length effects with
the multiple gluon emission using a poisson ansatz. The Higher-Twist
formalism given by Wang and Guo [24, 25] is another framework of the
parton energy loss. However, the obtained transport parameter from
the three set of parton energy loss calculations exists a
significant difference in magnitude. The similar discrepancies
between different parton energy loss formalisms are found in case of
heavy-ion collisions[32,33]. Therefore, the theoretical progresses
should be made furthermore. With future precision and much higher
energy semi-inclusive deep-inelastic scattering experiment and
theoretical advances in the parton energy loss, it should be
possible to achieve a truly quantitative understanding of the parton
energy loss mechanism.

\section{ Summary }

We study the hadron production from semi-inclusive lepton-nucleus
deep inelastic scattering, supplementing the perturbative QCD
factorized formalism with radiative parton energy loss. The
experimental data with the quark hadronization occurring outside the
nucleus are selected against the following criterion:  the hadron
formation time is more than twice the nuclear radius. The
fragmentation functions are modified with considering  the energy
loss incurred by hard quarks propagating through the nuclear medium
by means of SW  quenching weights and the analytic parameterizations
of quenching weights based on BDMPS formalism. The leading order
calculation of semi-inclusive hadron production on nuclei has been
done, and compared with the selected HERMES experimental data. In a
condition of considering the nuclear geometry effect on the
semi-inclusive hadron production, a good agreement between our
predictions and the experimental measurements is observed.  The
extracted  transport parameter from the global fit is found to be
$\hat{q} = 0.74\pm0.03$ GeV$^2$/fm for the SW  quenching weight
without the finite energy corrections. By using the analytic
parameterization of BDMPS quenching weight without the quark energy
$E$ dependence, the computed transport coefficient is $\hat{q} =
0.20\pm0.02$ GeV$^2$/fm.  It is found that the nuclear geometry
effect has a significant impact on the transport coefficient in cold
nuclear matter. Therefore, we should consider the nuclear geometry
effect in studying the semi-inclusive hadron production in deep
inelastic scattering on nuclear targets. In particular, we
emphasized that the semi-inclusive hadron production on nuclei may
help us to determine and constrain the transport coefficient (hence
the quark energy loss) in cold nuclear matter. Furthermore, a better
understanding of the parton energy loss mechanism can be achieved in
the future.

\vskip 1cm
{\bf Acknowledgments}
We are greatly indebted to the anonymous referees for insightful
suggestions that enabled us to develop a more comprehensive study.
Na Liu is grateful to Fran\c{c}ois Arleo for useful discussion. This
work was supported in part by the National Natural Science
Foundation of China(11075044, 11347107) and  Natural Science
Foundation of Hebei Province(A2008000137, A2013209299).

\end{document}